\documentclass[12pt]{article}

\usepackage{color,epsfig}

\begin{document}

\textwidth=14.8cm
\textheight=22.4cm
\oddsidemargin=-0.2cm
\topmargin=-1.6 cm

\def\spose#1{\hbox to 0pt{#1\hss}}\def\lta{\mathrel{\spose{\lower 3pt\hbox
{$\mathchar"218$}}\raise 2.0pt\hbox{$\mathchar"13C$}}}  \def\gta{\mathrel
{\spose{\lower 3pt\hbox{$\mathchar"218$}}\raise 2.0pt\hbox{$\mathchar"13E$}}}

\definecolor{vert}{rgb}{0.0,0.4,0.0}
\def\tt{{\color{blue}t}}  
\def\qq{{\color{red} {q}} }
\def\pp{{\color{vert} {\bf p}} }

\def\be{\begin{equation}}
\def\fe{\end{equation}}

\centerline{\bf Classical Anthropic Everett model: }
\centerline{\bf indeterminacy in a preordained multiverse.}
\medskip

\centerline{Brandon Carter}

\smallskip

\centerline{LuTh, Observatoire de Paris-Meudon.}

\centerline{Invited contribution for {\it Journal of Cosmology} {\bf 14}, 2011.}

\bigskip

\noindent

{\bf Abstract.} Although ultimately motivated by quantum theoretical 
considerations, Everett's many-world idea remains valid, as an approximation,
in the classical limit. However to be applicable it must in any case
be applied in conjunction with an appropriate anthropic principle,
whose precise formulation involves an anthropic quotient that can be 
normalised to unity for adult humans  but that would be lower for
infants and other animals. The outcome is a deterministic multiverse
in which the only function of chance is the specification of one's 
particular identity.

\bigskip

\noindent
{\bf 1. Introduction}
\medskip

Before the twentieth century, classical probabilistic models  -- such as those 
developed by Maxwell and Boltzmann for the treatment of many particle systems --
were commonly considered as approximations of an objective deterministic 
reality of which the details were unknown or at any rate too complicated to 
be tractable. However since the advent of quantum theory, it has come to be 
widely recognised that -- as Berkeley had warned -- such an objective 
material reality may not exist. A purported refutation  of the bishop's 
scepticism had been provided by Johnson's famous stone kicking experiment 
\cite{Deutsch97}, but the learned doctor might not have remained so
cockily confident if, instead of a tamely decoherent stone, he had
tried kicking the closed box containing Schroedinger's superposed 
live-and-dead cat. 

According to our modern understanding, classical probabilistic models should be 
considered as approximations, not of  illusory material reality,  but of more 
elaborate quantum theoretical models, whose interpretation is to a large extent 
subjective rather than objective. A complete understanding would therefor 
require a theory of the sentient mind -- as distinct from, though
correlated with, the physical brain.

The question of the relationship between our physical brains -- the object of 
study by neurologists -- and the thoughts and feelings in our ``conscious'' 
minds was already  a subject of philosophical speculation long before 
the development of quantum theory. As very little substantial progress
had  been achieved, it was natural that some people should wonder whether a 
resolution of the mystery of quantum theory might provide a resolution
of the mystery of the mind. A more common opinion has however been
expressed by Steven Weinberg, who wrote\cite{Weinberg} ``Of course 
everything is ultimately quantum mechanical: the question is whether 
quantum mechanics will appear directly in the theory of the mind, 
and not just in the deeper level theories like chemistry on which 
the theories of the mind will be based ... Penrose
may be right about that, but I doubt it.''

I am inclined to share this common opinion, and will proceed here on
that basis, not just because of the relatively macroscopic (multiparticle) 
nature of the neurons constituting the brain, but because quantum
theory is not really essential for what is commonly considered to be
the crux of the mind to matter relationship, namely what is known as
the ``collapse'' of the ``wave function'' which is supposed to result
from an observation of the kind exemplified by Schroedinger's gedanken 
experiment in which a cat in a box is liable to be killed by a pistol 
triggered by a Geiger counter.

\bigskip
{\bf 2. The trouble with the traditional doctrine}
\medskip

According to the ``Copenhagen'' interpretation, the relevant ``wave function''
collapses either to a pure state in which the  cat is unambiguously alive, or 
else one in which it is unambigously dead, when a human ``observer''opens its 
box. The trouble with the Copenhagen interpretation is that it denies 
``observer'' status to the occupant of the box, which is questionable even in 
the case of a humble cat, and would clearly be quite inadmissible if
the cat were replaced by another human. 

However as well as the underlying symmetry between the person at risk and the
person who observes, the point I want to emphasize here is that the issue is 
not essentially quantum mechanical, because it subsists even if one goes over 
to the (decoherent) classical limit. In the human case, an analogous classical
experiment can be -- and historically has been --  done with the Geiger 
triggering mechanism replaced by use of  an old fashionned Russian roulette 
revolver. The classical analogue of the ``collapse of the wave function'' 
would be the Bayesian reduction of the corresponding classical probability 
distribution, from an  {\it a priori} configuration, in which the outcome is 
uncertain, to an {\it a posteriori} configuration in which the  subject of the 
experiment is either unambiguously alive or else unambiguously dead.

To the question of which protagonist has the privilege of making the 
observation whereby the definitive ``collapse'' occurs, it is
traditionally presumed that Bishop Berkeley's reply would have been
been ``God!''. However physicists (since the time of Laplace) have
tried to avoid such ad hoc invocation of a ``deus ex machina'', and (in 
the spirit of Ockham's razor) are therefore inclined to prefer 
the alternative reply that is expressible succinctly as
``None!''. Such negation was originally proposed by Everett, and was 
advocated -- but not adequately elucidated -- first by Wheeler and 
subsequently by DeWitt \cite{DeWitt}. By thus denying the Copenhagen 
doctrine of the occurrence of ``collapse'' as an objective physical
process -- rather than merely a subjective allowance for new information 
as in the familiar classical case of Bayesian reduction -- Everett got
off to a good start. However his attempt to provide a positive
interpretation of the meaning of the ``wave function'' was not
entirely successful. 

Part of the trouble arose merely from misunderstanding, due to injudicious 
choice of wording, whereby what I would prefer to refer to as alternative 
``channels'' were called ``branches'', thereby conveying the misleading 
idea of a continual multiplication of worlds \cite{Leslie96}, whereas 
(since Everett's idea was that evolution remains strictly unitary) the 
``worlds'' in question are strictly conserved, having neither begining
nor end: what changes is only the resolution of distinction between
different ``channels'',  which may become finer (or coarser!) as 
observational information is acquired (or lost!). A more serious - since 
not merely semantic -- problem by which many people have been puzzled
is what Graham\cite{Graham} has called the ``dilemma'' posed by Everett's 
declaration that the alternative possible outcomes of an observation are 
all ``equally real'' though not (if their quantum amplitudes are different) 
``equally probable''. As I have argued previously \cite{C04}, and will 
maintain here, the resolution of this dilemma requires the invocation
of an appropriate anthropic principle.

\bigskip
{\bf 3. The concept of reality}
\medskip

It was recognised long ago by Berkeley, and has been emphasized more
recently by Page \cite{Page96}, that the only kinds of entities we know 
for sure to be real are our mental feelings and perceptions (including 
dreams).  The material world in which we have the impression of living
is essentially just a theoretical construct to account for our perceptions. 
In the dualist (Cartesian) picture that used to be widely accepted,
this material world was supposed to have a reality of its own, on  par with 
the realm of feelings and perceptions. However under closer scrutinary 
such separate material reality has turned out to be illusive, so we find 
ourselves glimpsing a more mysterious but apparently unified quantum picture. 
Following the approach initiated by Everett \cite{DeWitt,Graham}, diverse 
attempts to sketch the outlines of such a unified picture have been made, 
albeit with only rather limited success so far, by various people 
\cite{Deutsch99,Wallace03,Greaves04}, and in particular -- from a
point of view closer to that adopted here -- by the present author 
\cite{C04}, and by Page \cite{Page96}. 

Assuming, as remarked above, that mental processes have an essentially 
classical rather than quantum nature, this essay has the relatively modest 
purpose of attempting to sketch the outlines of a simpler, more easily 
accessible,  classical unification that may be useful pedagogically and, 
in appropriate circumstances, as an approximation to a more fundamental 
quantum unification that remains elusive. The picture proposed here is
based on the use of an appropriate anthropic principle in conjunction 
with the Everett approach, which is relatively well defined in the 
classical limit, so that the notions of ``equal reality'' and ``unequal 
probability'' can be clarified in a coherent manner. 

Deutsch, Wallace, and Greaves have developed an alternative approach 
\cite{Deutsch99,Wallace03,Greaves04} that attempts to do this in terms
of the kind of probability postulated in decision theory, on the
debatable supposition that the relevant observations are performed by 
``rational agents''. 

The essentially different approach advocated here is based on probability 
of a kind proportional to the amount of perception that is ``real'', in the 
sense not of Deutsch \cite{Deutsch97} but of Page\cite{Page96} -- as based 
on sentience rather than rationality.   Following a line of thought
originated by Dyson \cite{Dyson79}, I have suggested \cite{C04,C05} that the
relevant amount of perception should in principle be measured by the 
corresponding Shannon type information content, but in practice that does not 
tell us much, as it leaves us with the unsolved  question of which of the 
many processes going on in the brain are the ones that actually correspond 
to sentient perception. This fundamental question does not matter so long as 
we are concerned only with the standard, narrowly anthropic, case of adult 
humans, for whom (as in the example of the next section) it can 
reasonably be assumed that such processes go on at roughly the same average 
rate. However for more general applications it would be necessary to face the 
intractible problem of estimating the relevant anthropic quotient $\qq$, 
meaning an appropriate correction factor that might be larger than unity for 
conceivable extraterrestrials, but that would presumably be smaller for 
extinct hominids, and much smaller for other animals as well as for infants 
of our own species. The easiest non-trivial case to deal with would presumably 
be that of ordinarily senile members of our own species, as their mental 
processes are similar to those of adults in their prime except for a reduction 
in speed that can be allowed for by a factor $\qq$ that should be clinically 
measurable (and of practical interest for therapeutic purposes).

\bigskip
{\bf 4. Russian roulette: a historical example}
\medskip

It is customary \cite{Greaves04} to demonstrate the application of such 
ideas by idealised gedanken experiments in which, if there are just two 
protagonists, their initials are commonly taken to be A for Alice and
B for Bob (while to illustrate merely logical, rather than physically 
conceivable possibilities, it is common \cite{Bostrom} to consider examples 
that are not just idealised but frankly fantastic, in which case the 
protagonists are referred to as ``Adam and Eve''). However to emphasise 
that I am concerned with what is ``real'' I shall take as a (simplified 
and approximate, but not artificially idealised) example an experiment that 
is not merely hypothetical, but that really occured as a historical
event during the XXth century, with a principle protagonist whose initial 
was actually not A but G.

To illustrate the basic idea, I propose to consider a modified Schroedinger 
type experiment in which G  -- an unbalanced adolescent at the time -- 
voluntarily and crazily took the role of the cat, in a solitary game of 
Russion roulette. The role of the external observer was taken by his big 
brother (the owner of the revolver) to whom I shall refer by the letter B. 
Having first heard about it privately from someone who had been neighbour at 
the time, I read about it many years later in published memoirs of G, who 
not only survived the experiment but recovered his mental equilibrium and 
lived to a ripe old age -- at least in our particular branch-channel of the 
multiverse.

To keep the arithmetic simple, I shall postulate that the revolver was just a 
five-shooter, of the compact kind that is most convenient as a concealed 
weapon. (In reality it may well have been a six-shooter of the kind familar in 
cowboy movies, but it can safely be presumed that it was not what was 
originally used by the reputed inventors of Russian roulette, namely Czarist 
officers, whose standard service revolvers were actually seven-shooters.) The 
protocol of the potentialy suicidal game is to load just one of the cartridge 
chambers and then to whirl it to a random position before pulling the trigger. 
In such a case, starting from initial conditions that are imperceptibly 
different, there will be five equally likely outcomes, of which four will be 
indistinguishable for practical purposes, whereas the other one will be fatal.

According to the traditional single-world doctrine of deterministic classical 
physics (as still taken for granted at the time of the incident in question) 
only one of the five possible outcomes would have actually occurred. However 
according to the Everett type many-world doctrine, a complete description
 will involve many separately conserved ``strands'' (commonly but misleadingly 
referred to as ``branches'') meaning single worlds of all the five types, in a 
multiverse consisting of five equally numerous sub-ensembles or ``channels'', 
one for each qualitatively distinct possibility. Such sub-ensembles will be 
characterised by a physical measure given by the fraction $\pp$ of the total 
number of strands, which in this case is $\pp=1/5$ for each one. Since the four 
possibilities in which G survives would have been effectively distinguishable 
(by an examination of the weapon) only for a very short time after the 
experiment, it will in practice be sufficient  for most subsequent purposes
to use a coarser representation in which they are regrouped into a single 
larger multistrand ``channel'', which will thus have  measure $\pp=4/5$. When 
Everett refers to things as ``equally real'' it is clear that he should be 
understood to have in mind the individual (single world) strands, rather than 
their weighted groupings into broader ``channels''. 

The stage at which the original presentation of the Everett approach  becomes 
unclear is when it is suggested that the physical weighting introduced as 
described should somehow be interpreted as a probability, despite the fact 
that (as the classical limit of evolution that is strictly unitary in 
the quantum case) the behaviour of the many worlds involved is entirely 
deterministic, so that when their initial configurations have been specified 
no uncertainties remain. 

To give a meaning to the concept of probability in this context, the purely 
materialistic framework of the classical many-world system described so far 
needs to be extended to include allowance for the role of mind. For the 
simplified classical model considered here, it will be good enough to do this 
in the usual way, by supposing that mental feelings and perceptions correspond 
to physical states of animate brains that are roughly localisable on time 
parametrised world lines of the animals concerned within the single world 
``strands''. 

\bigskip
{\bf 5. Anthropic quotient}
\medskip

Within the foregoing framework, the incorporation of probability into the model 
is  achieved by an appropriate application of the anthropic principle. In the 
simple (weak) version that is adequate for the present purpose, the anthropic 
principle \cite{C05,C10} prescribes that the probability of finding oneself 
on a particular animate world line on a single strand within a small time 
interval ${\rm d}\tt$ is proportional to $\qq\, {\rm d}\tt$, where the 
``anthropic quotient'',  $\qq$, is normalised to unity in the average (adult) 
human case. This coefficient $\qq$ is interpretable as a measure of the 
relative rate of conscious sentient thought (which might be very low compared
with the rate of subconscious but perhaps highly intelligent information 
processing, such as could be performed by an insentient computer). Whereas 
it might be higher than unity for conceivable extraterrestrials,  $\qq$  
would presumably be lower for other terrestrial species (such as chimpanzees) 
as well as for infants and senile members of our own species (see Figure 
\ref{refinedbiograph}). On short (diurnal) timescales the anthropic quotient 
of an individual would fluctate between high waking levels and low dreaming 
values, and it would of course go to zero at and after the instant of death, 
as also before conception (though perhaps not before the instant of birth).

In the CAE (classical anthropic Everett) model set up in this way, the meaning 
of the weighting fraction $\pp$ of a channel constituted by an ensemble of very 
similar single-world strands is now clear. It does not directly determine the 
total probability of finding oneself in that channel, but it does determine the
probability ${\rm d}P$ of finding oneself within a time interval ${\rm  d}\tt$ 
on a world line of a particular kind (such as that of G, oralternatively 
that of B in the example described above) within the channel in question,
according to the specification ${\rm d}P\propto \pp\, \qq\, {\rm  d}\tt$ 
(with the proportionality factor adjusted so that the total
probability for all possibilities adds up to unity).

\begin{figure}
\centering
\epsfig{figure=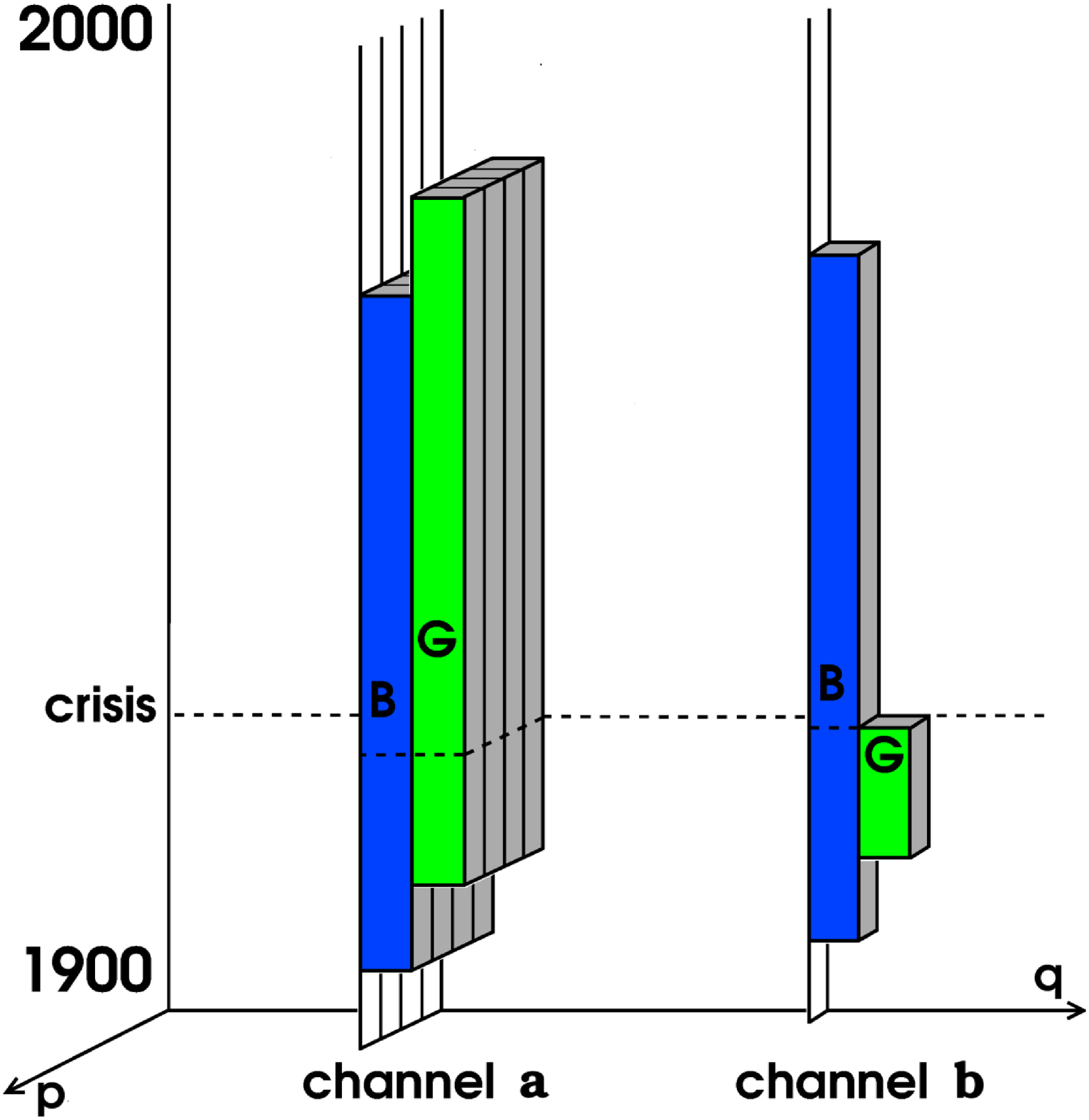, width=12.0cm, 
}
\caption{Crude anthropic biograph of XXth century roulette gamester G 
(pale shading) and his brother B (dark shading) using the vertical direction 
for time, while the thickness of a worldline in the sideways direction  
measures subjective anthropic probability weighting per unit time, as 
specified by the anthropic quotient $q$ which is set to zero before birth 
and after death, and is taken here to have a uniform unit (average) value 
during life -- whereas in a less crude version it would taper off at the 
beginning (infancy) and the end (senility). For each of the (Everett type) 
channels, the 3rd dimension -- out of the page or screen --  measures the 
number of ``strands'', representing the objective physical probability $p$ 
(the square of a corresponding quantum amplitude) which is conserved. In 
such an anthropic diagram, the probability  of finding oneself to be in a 
particular state of a particular person during a particular time interval 
is proportional to the relevant volume (the time integral of the product 
$pq$ of the anthropic and physical  probability measures). In channel {\bf a}
 -- the one we know about historically, because we are on it ourselves -- 
both brothers survived through a complete life span until old age, the 
younger naturally outliving the elder. In channel {\bf b}, for which  the 
life of G was truncated after only one 6th of its natural span,  it is 
supposed that the subsequent life of B would not have been substantially 
affected.
}
\label{crudebiograph}
\end{figure}

\begin{figure}
\centering
\epsfig{figure=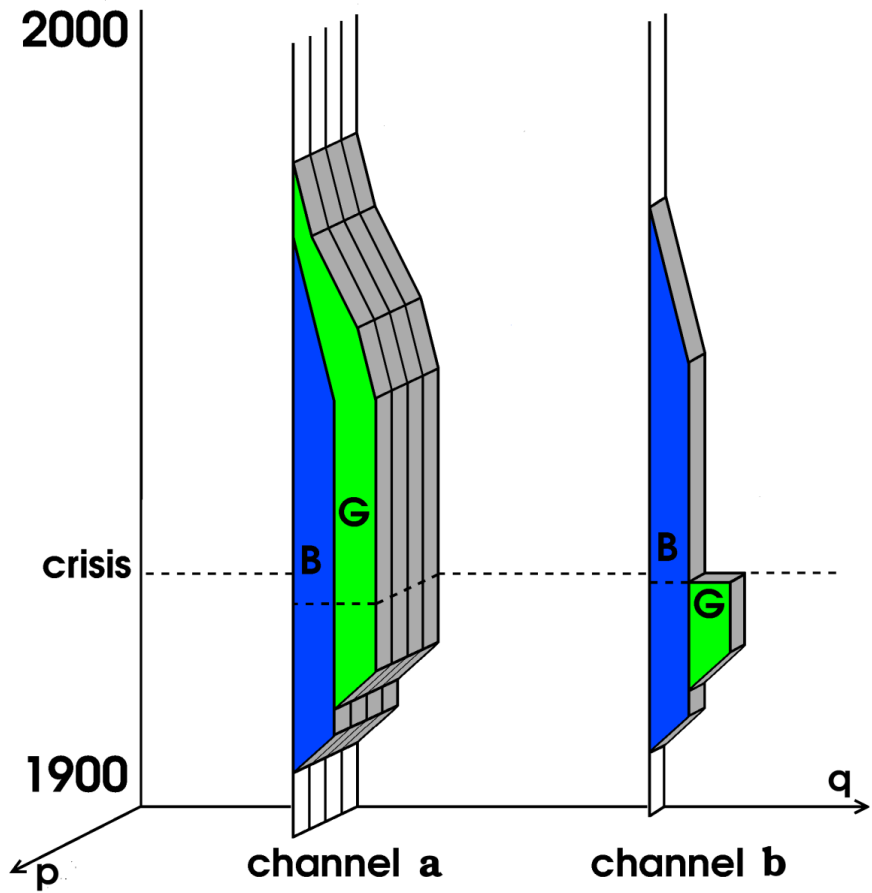, width=12.0 cm, 
}\caption
{Refined anthropic biograph for  Russian roulette gamester G and 
his brother B, using same conventions as in Figure \ref{crudebiograph}, 
but allowing for non uniformity of the anthropic quotient $q$, which 
(for each protagonist) is taken to have a piecewise linear dependence 
on age, rising rapidly in infancy, and declining more slowly as
senility sets in.
\label{refinedbiograph}
}
\end{figure}

Let us see how this works out in the simple example of the roulette
gamester G and his brother B, as shown in Figure \ref{crudebiograph}, on the 
assumption that both can be considered as average adults characterised by
$\qq=1$. To keep the figures round, let us take it that in the first
channel, labelled {\bf a} , with $\pp_{\rm \bf a}=4/5$, both roulette
gamester G and his brother B survived 6 times longer (to an age of
about 90) than G did in the second (fatal) channel, labelled {\bf b},
with $\pp_{\rm \bf b}=1/5$, where the life of B would have been unaffected 
(while that of G would have been truncated at about age 15). This can
be seen to imply that one is 20 per cent more likely to find oneself to 
be B than G. In the former case, one will have a 20 per cent chance of being 
in the fatal channel, and thus of witnessing the death of one's younger 
brother. In the latter case, that is to say conditional on being G, one will 
have a 20 per cent chance of finding oneself in the time interval before the 
game, and thus with only a 4 per cent chance of being in what will
turn out to be the fatal channel.

If B and G had been the only sentient inhabitants of the world it can be seen
that the {\it a priori} odds against channel {\bf b} would have been  48 to
7 which is almost 7 to 1. However when account is taken of all the rest 
of the population (who would not have been significantly affected by the 
outcome of the roulette game) it can be seen that the {\it a priori} odds 
against finding oneself in channel {\bf b} (and thus deprived of access to  
G's later literary output) would actually have been barely greater than 4 
to 1 (the value given by the ratio $\pp_{\rm \bf a}/\pp_{\rm \bf b}$ of the 
naive physical probabilities designated by $\pp$).

A more complete picture, allowing for the many other inhabitants of the world, 
would of course require a much finer decomposition involving far more than two 
qualitatively distinct channels. Indeed a complete multibiography just of G 
alone would probably require many more channels to allow for the
vicissitudes of his later life, which extended not just through the
Second World War but even through the Cold War. In particular -- to be
fully realistic -- an adequate multi-history of the latter would
presumably require the inclusion of non-neglibly weighted channels in
which an incident such as the Cuban missile crisis terminated in the 
disastrous manner envisaged by Shute \cite{Shute}.

\bigskip
{\bf 6. Commentary}
\medskip

Although the interpretations -- and perhaps the ethical implications -- are 
different, there is no effectively observable distinction between what is 
predicted by the deterministic many-world CAE model presented here (in which 
only one's identity is unforeseeable) and what is predicted by the 
corresponding classical model of the ordinary single-world type (in which the 
material physical outcome depends on chance). It might therefore be argued 
that the traditional single-world model should be preferred on the
grounds that it is simpler, or less ontologically ``bloated''. It is
however to be recalled that a classical model cannot claim to
represent ultimate reality, but merely provides what is at best an 
approximation to a more accurately realistic quantum model, a purpose
for which the traditional single world-model is not so satisfactory. 

Another point to be emhasised is that the ontology in question involves only 
mental feelings and perceptions. As foretold by Berkeley, but contrary to what 
used to be taught by ``positivists'' such as Mach, matter, as incorporated in 
physical fields over spacetime, should not be considered to have  objective 
``reality'', but has the status merely of mathematical machinery (that might 
be replaced for predictive purposes by an equivalent action at a distance 
formulation based on Green functions).

Having recognised that the relevant ontology does not involve matter but only 
mind, one is still free to entertain different opinions about how extensive or 
``bloated'' \cite{Leslie83} it may be.  The anthropic measure characterised by 
the coefficient $q$ merely determines the relative probability of the 
perceptions in question, but but not the absolute number of times they occur. 
If ontological economy is a desideratum, it might seem preferable to postulate 
the actual occurrence only of a fraction of the perceptions admitted by the 
theory. On the other hand for those concerned with economy only in the sense 
of Ockham's razor, and particularly for those who are unhappy with the
concept of probability except when it can be prescribed in terms of relative 
frequencies, the most attractive possibility would presumably be to suppose 
that all the perceptions admitted by the theory actually occur (in the 
indicated proportions). Although it is more ontologically extravagant, the
latter alternative has the advantage of conforming to the requirement that 
was expressed in metaphorical language by Einstein's edict that ``God does
not play dice''. However that is not for us mere mortals to judge: as far as 
scientific observation is concerned there is no way of telling the difference.

A more mundane issue (with ethical implications concerning protection from 
inhumane treatment) is the evaluation of the appropriate anthropic quotient 
$\qq$ for non-human terrestrial animals (such as the cat considered by 
Schroedinger) and particularly for infants of our own species. It is to 
be presumed that $\qq$ should be of the order of unity for extinct hominids 
such as {\it homo erectus}, whose integrated population time is at most 
comparable with our own \cite{C10}. However the observation that we do
not belong to the far more numerous populations of animals of other,
less closely related, kinds suggests that their anthropic quotients
should be much lower, and hence, by analogy, that the same may apply to 
infants.

\bigskip
{\bf Acknowledgement}
\medskip

The author 
is grateful for stimulating discussions with John Wheeler, Roger Penrose, 
and Bryce DeWitt on various past occasions, and also for more recent
discussions with John Leslie and Don Page. 

\vfill\eject

\end{document}